\documentclass[11pt,a4paper]{article}
\pdfoutput=1
\usepackage{jheppub}
\usepackage{cite}

\usepackage{wrapfig,boxedminipage,subfigure,epsfig}
\usepackage{amsxtra,amstext,latexsym,dsfont,amsfonts}
\usepackage{color}
\usepackage[dvipsnames]{xcolor}
\usepackage{float}
\usepackage{slashed}
\usepackage{cancel}
\usepackage{calligra}
\usepackage{graphicx,amsmath,amssymb,amsthm,multirow,array,bm,bbm,esint}
\usepackage{tikz} 
\usepackage{array,tikz-cd}
\usetikzlibrary{decorations.pathmorphing}
\DeclareFontShape{T1}{calligra}{m}{n}{<->s*[2.2]callig15}{}
\DeclareMathAlphabet{\mathcalligra}{T1}{calligra}{m}{n}

\newcommand{\be}{\begin{equation}}
\newcommand{\ee}{\end{equation}}
\newcommand{\bea}{\begin{eqnarray}}
\newcommand{\eea}{\end{eqnarray}}

\newcommand{\dd}{\mathrm{d}}

\title{Holographic scalar and vector exchange in OTOCs and  pole-skipping phenomena}

\author[]{Keun-Young Kim,}
\author[]{Kyung-Sun Lee,}
\author[]{and Mitsuhiro Nishida}

\affiliation[]{School of Physics and Chemistry, Gwangju Institute of Science and Technology, 123 Cheomdan-gwagiro, Gwangju 61005, Korea}

\emailAdd{fortoe@gist.ac.kr}
\emailAdd{kyungsun.cogito.lee@gmail.com }
\emailAdd{mnishida@gist.ac.kr}

\abstract{We study scalar and vector exchange terms in out-of-time-order correlators (OTOCs)  holographically. By applying a computational method in graviton exchange, we analyze exponential behaviors in scalar and vector exchange terms at late times. We show that their exponential behaviors in simple holographic models are related to pole-skipping points obtained from the near-horizon equations of motion of scalar and the vector fields. 
Our results are generalizations of the relation between the graviton exchange effect in OTOCs and the pole-skipping phenomena of the dual operator, to scalar and the vector fields. 
 }

\begin{document}
\maketitle



\section{Introduction}

Quantum field theories (QFTs) that have the gravity duals are in a special class of QFTs, and their properties are studied using various field theoretical and holographic methods.
An interesting property of quantum systems that are described by black holes is the saturation of the quantum Lyapunov exponent $\lambda_L=2\pi /\beta$ \cite{Shenker:2013pqa, Roberts:2014isa, Roberts:2014ifa, Shenker:2014cwa, Kitaev-2014, Perlmutter:2016pkf}, where $\beta$ is inverse temperature. The  Lyapunov exponent $\lambda_L$ in four-point out-of-time-ordered correlators (OTOCs) is a diagnosis of quantum chaos \cite{Kitaev-2014, larkin1969quasiclassical}. It has been proposed that $\lambda_L$ in quantum many-body systems is bounded by $\lambda_L\le2\pi /\beta$ \cite{Maldacena:2015waa} based on reasonable physical assumptions .

It was found that the retarded Green's function of energy density in momentum space contains information on quantum chaos in the quantum systems with the gravity duals \cite{Grozdanov:2017ajz}, which supports a connection between transport properties and quantum chaos. In particular, a pole-skipping point $(\omega_*, k_*)$ in the retarded Green's function of energy density would be related to the Lyapunov exponent $\lambda_L$ and butterfly velocity $v_B$ as follows:
\begin{align}
\omega_*=i\lambda_L, \;\; k_*=i\lambda_L/v_B.
\end{align}
The pole-skipping points of a Green's function $G(\omega, k)$ are defined as the intersection points between lines of poles and lines of zeros in $G(\omega, k)$. This relation is called ``pole-skipping phenomena" \cite{Grozdanov:2017ajz, Blake:2017ris}\footnote{The pole-skipping phenomena in non-maximally chaotic theories have been proposed in \cite{Choi:2020tdj} as follows: If the energy-momentum tensor exchange dominates in a certain range, a pole-skipping point of the energy density Green's function encodes the energy-momentum tensor contribution to quantum chaos.}. To compute the pole-skipping points holographically, a near-horizon analysis  has been formulated \cite{Blake:2018leo}. By examining special points in the equations of motion (e.o.m.) near the black hole horizon, one can obtain the pole-skipping points.

After the discovery of the pole-skipping phenomena of energy density, pole-skipping points of other fields such as scalar, vector, and spinor have been investigated holographically \cite{Grozdanov:2019uhi, Blake:2019otz, Natsuume:2019xcy, Natsuume:2019vcv, Wu:2019esr, Ceplak:2019ymw, Abbasi:2019rhy,  Ahn:2020bks, Ahn:2020baf, Natsuume:2020snz}.\footnote{For other progress in holographic studies of pole-skipping points, see for example, \cite{Grozdanov:2018kkt,Li:2019bgc,Natsuume:2019sfp, Ahn:2019rnq,  Liu:2020yaf, Abbasi:2020ykq, Jansen:2020hfd, Grozdanov:2020koi}. } However, the pole-skipping points of the lower spin fields $(\ell\le1)$ do not exist in the upper-half plane of the complex $\omega$-space, and quantum chaos does not seem to be related to their pole-skipping points. This observation can be interpreted in terms of the holographic computation method for the Lyapunov exponent and butterfly velocity. In holographic models, they can be computed from shock wave geometry which is made with graviton exchange (see, for example, \cite{Blake:2016wvh, Roberts:2016wdl, Shenker:2013pqa, Roberts:2014isa, Shenker:2014cwa, Ahn:2019rnq}). Because the energy-momentum tensor corresponds to the graviton in holography, the pole-skipping point of energy density would be related to the graviton exchange term in the OTOCs, which is relevant to quantum chaos in the holographic models. By extending this interpretation to other fields, one can expect the pole-skipping points of them to be related to ``exchange terms" other than the graviton exchange in the OTOCs.

In conformal field theories (CFTs), these exchange terms correspond to conformal blocks with an analytic continuation for the OTOCs. The pole-skipping points of scalar, vector, and energy density in CFTs on hyperbolic space $\mathbb{H}^d$ with $\beta=2\pi$ were computed by \cite{Haehl:2019eae, Ahn:2020bks},\footnote{The pole-skipping points in CFTs were also studied in \cite{Liu:2020yaf, Haehl:2018izb, Das:2019tga, Ramirez:2020qer}.} and it was shown that they are related to late time exponential behaviors of the conformal blocks. This result matches with the previous expectation and is regarded as a generalization of pole-skipping phenomena to other fields, although the pole-skipping points of scalar and the vector fields are not related to maximal chaos.

From the holographic viewpoint, one can compute the pole-skipping points of CFTs on $\mathbb{H}^d$ with $\beta=2\pi$ by using a $(d+2)$-dimensional AdS-Rindler black hole geometry \cite{Ahn:2019rnq, Ahn:2020bks}. To study the  generalization of pole-skipping phenomena holographically, it is useful to develop a holographic computation method of exponential behaviors in exchange terms other than the graviton exchange. In particular, computations in a planar AdS black hole are important because the pole-skipping points on flat space $\mathbb{R}^d$ are well-studied compared to the ones on $\mathbb{H}^d$.

In this paper, we study the exponential behaviors of scalar and vector exchange terms in the four-point OTOCs by using the holographic method. We use simple holographic models, which have three-point interactions to compute the exponential behaviors. Our computation method is a generalization of computations for the Lyapunov exponent and butterfly velocity from the graviton exchange term. By comparing them with the near-horizon analysis, we check that the exponential behaviors are related to the pole-skipping points in the retarded Green's function of scalar and the vector fields.

The paper is organized as follows. The calculation of exponential behavior in the graviton exchange term is reviewed in Section \ref{sec2}. In Sections \ref{sec3} and \ref{sec4}, we compute exponential behaviors in the scalar and vector exchange terms and compare them with the pole-skipping points derived from the near-horizon analysis. We discuss our conclusion and future work in Section \ref{sec5}.

\section{Review: exponential behavior with graviton exchange}\label{sec2}
We review the calculation of exponential behaviors in the graviton exchange term based on \cite{Blake:2016wvh, Roberts:2016wdl}. From this calculation, we can obtain the Lyapunov exponent and butterfly velocity in the holographic systems. In the subsequent sections, we will generalize this computation for the scalar and vector exchange terms.

For a holographic computation of OTOC $\langle W(t_W, \mathbf{x}_W)V(0,\mathbf{x}_V)W(t_W, \mathbf{x}_W)V(0,\mathbf{x}_V)\rangle$, we consider the Einstein-Hilbert action and the scalar fields actions: 
\begin{align}
S&=S_W+S_V  + S_{\mathrm{EH}} \,,\\
S_W&=-\frac{1}{2}\int d^{d+2}x \sqrt{-g} \left(g^{\mu \nu} \partial_{\mu} \phi_W \partial_{\nu} \phi_W+m_W^2 \phi_W^2 \right)\,,\label{SW}\\
S_V&=-\frac{1}{2}\int d^{d+2}x \sqrt{-g} \left(g^{\mu \nu} \partial_{\mu} \phi_V \partial_{\nu} \phi_V+m_V^2 \phi_V^2 \right)\,,\label{SV}\\
S_{\mathrm{EH}}&=\int d^{d+2}x \sqrt{-g}\left(R-2\Lambda\right),
\end{align}
where $\Lambda$ is the cosmological constant. These actions determine bulk propagators of the scalar fields and gravitons. The bulk scalar fields $\phi_W$ and $\phi_V$ correspond to the boundary operators $W$ and $V$ in the four-point OTOC $\langle W(t_W, \mathbf{x}_W)V(0,\mathbf{x}_V)W(t_W, \mathbf{x}_W)V(0,\mathbf{x}_V)\rangle$. To compute exponential behaviors in the OTOC holographically, we assume that $W$ is a heavy operator and treat $W(t_W, \mathbf{x}_W)$ as a source as in \cite{Afkhami-Jeddi:2017rmx}. Because $\phi_W$ is coupled with graviton as shown in (\ref{SW}), the source $W(t_W, \mathbf{x}_W)$ makes a shock wave geometry \cite{Aichelburg:1970dh, Dray:1984ha, Sfetsos:1994xa} on the bulk side. 

\begin{figure}
\centering

\begin{tikzpicture}[scale=1.5]
\draw [thick]  (0,0) -- (0,3);
\draw [thick]  (3,0) -- (3,3);
\draw [thick,dashed]  (0,0) -- (3,3);
\draw [thick,dashed]  (0,3) -- (3,0);
\draw [thick,decorate,decoration={zigzag,segment length=1.5mm, amplitude=0.3mm}] (0,3) .. controls (.75,2.85) 
and (2.25,2.85) .. (3,3);
\draw [thick,decorate,decoration={zigzag,segment length=1.5mm,amplitude=.3mm}]  (0,0) .. controls (.75,.15) and (2.25,.15) .. (3,0);

\draw[thick,<->] (1,2.2) -- (1.5,1.7) -- (2,2.2);
\fill (0,0.2) circle [radius=2pt];
\fill (3,2.8) circle [radius=2pt];
\end{tikzpicture}
\vspace{0.1cm}
\put(-195,8){$W(t_W, \mathbf{x}_W)$}
\put(-0,117){$W(t_W, \mathbf{x}_W)$}
\put(-98,98){\Large $\tilde{u}$}
\put(-48,96){\Large $\tilde{v}$}
 \caption[A]{Penrose diagram of the AdS black hole space-time at $\mathbf{x}=\mathbf{x}_W$ with an appropriate transformation $u\to\tilde{u}(u)$ and $v\to\tilde{v}(v)$. See \cite{Ahn:2019rnq} for more details and the definition of $\tilde{v}$ and $\tilde{v}$. Wavy lines are the singularities, dashed lines are the horizons at $\tilde{u}=u=0$ and $\tilde{v}=v=0$, and vertical lines are the AdS boundaries. In the holographic computations of OTOCs,  we usually insert two operators $W(t_W, \mathbf{x}_W)$, whose Euclidean times differ by $\beta/2$, at the left and right boundaries.}\label{penrose}
\end{figure}
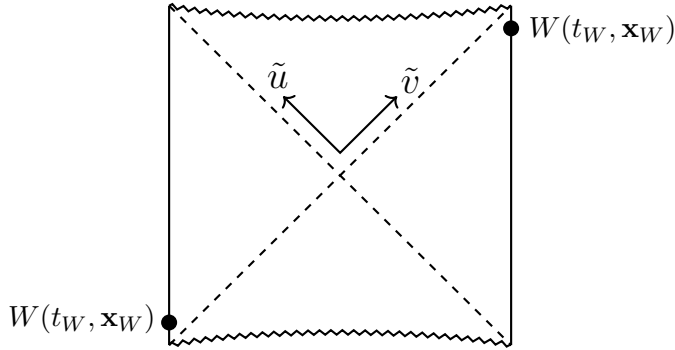

As an initial metric before making the shock wave geometry, consider a black hole metric\footnote{To obtain a rich variety of background metrics, one can consider extra matter fields in addition to $S_{\mathrm{EH}}$. See \cite{Sfetsos:1994xa} for a detailed analysis of the shock wave geometry with matter fields.} 
\begin{align}
\dd s^2&= -U(r)\dd t^2+\frac{\dd r^2}{U(r)}+V(r)\dd \mathbf{x}^2,\label{osbhm}
\end{align}
where $\dd \mathbf{x}^2$ is the squared line element of boundary space $M$ which does not have a periodic direction. In this paper we mainly focus on $M=\mathbb{R}^d$ and $M=\mathbb{H}^d$. The Hawking temperature of this black hole is $T=1/\beta=U'(r_0)/4\pi$, where $r_0$ is the horizon radius. By using Kruskal coordinates $(u,v)$
\begin{align}
uv=-e^{U'(r_0)r_*(r)}, \qquad u/v=-e^{-U'(r_0)t}, \qquad dr_*=dr/U(r),\label{Kc}
\end{align}
we can extend (\ref{osbhm}) to a two-sided black hole metric
\begin{align}
\dd s^2&=2A(uv)\dd u\dd v+B(uv)\dd \mathbf{x}^2, \label{tsbhm}\\
A(uv)&=\frac{2}{uv}\frac{U(r)}{U'(r_0)^2}, \qquad B(uv)=V(r).\notag
\end{align}
As an example, the Penrose diagram of the AdS black hole space-time is shown in Fig.~\ref{penrose}.

At late times $t_W\gg\beta$, a geodesic between the boundaries of the two-sided black hole on the $t=t_W$ slice approaches  the horizon $u=0$ and an expectation value of the energy-momentum tensor $T_{\mu\nu}=-\frac{2}{\sqrt{-g}}\frac{\delta S_{W}}{\delta g^{\mu\nu}}$ is localized on the horizon $u=0$  \cite{Aichelburg:1970dh, Dray:1984ha, Sfetsos:1994xa, Shenker:2013pqa, Roberts:2014isa, Shenker:2014cwa}:
\begin{align}
\frac{\langle\psi|T_{uu}(u, v, \mathbf{x})|\psi\rangle}{\langle\psi|\psi\rangle}=Pe^{\frac{2\pi}{\beta}  t_W}\delta (u)\delta (\mathbf{x}-\mathbf{x}_W)\,,\label{evemt}
\end{align}
where $|\psi\rangle$ is a dual state of the two-sided black hole with the source $W(t_W+i\tau, \mathbf{x}_W)$ \cite{Shenker:2013pqa, Roberts:2014ifa}, and $P$ is related to the initial asymptotic momentum of the source.  Now, we introduce Euclidean time $\tau$ for regularization. As we will see in the subsequent sections, the $t_W$-dependence $e^{\frac{2\pi}{\beta} t_W}$ in (\ref{evemt}) is related to the spin $\ell$ of exchange fields as $e^{\frac{2\pi}{\beta} (\ell-1) t_W}$, and it is consistent with the late time behavior of the  conformal block \cite{Roberts:2014ifa, Perlmutter:2016pkf}.

If we assume holographic correspondence, correlation functions of QFTs can be computed from bulk scattering amplitude \cite{Maldacena:1997re, Gubser:1998bc, Witten:1998qj}. At the late times $t_W\gg\beta$, the momentum of particles around the horizon becomes exponentially large, as seen from (\ref{evemt}), and the bulk scattering is regarded as high-energy scattering. Therefore, one can use the eikonal approximation at the late times. In the eikonal approximation with a large distance limit, massless graviton $(\ell=2)$ exchange is dominant in holographic models \cite{Levy:1969cr, tHooft:1987vrq, Cornalba:2006xk}.

\begin{figure}
\centering
     {\includegraphics[width=5cm]{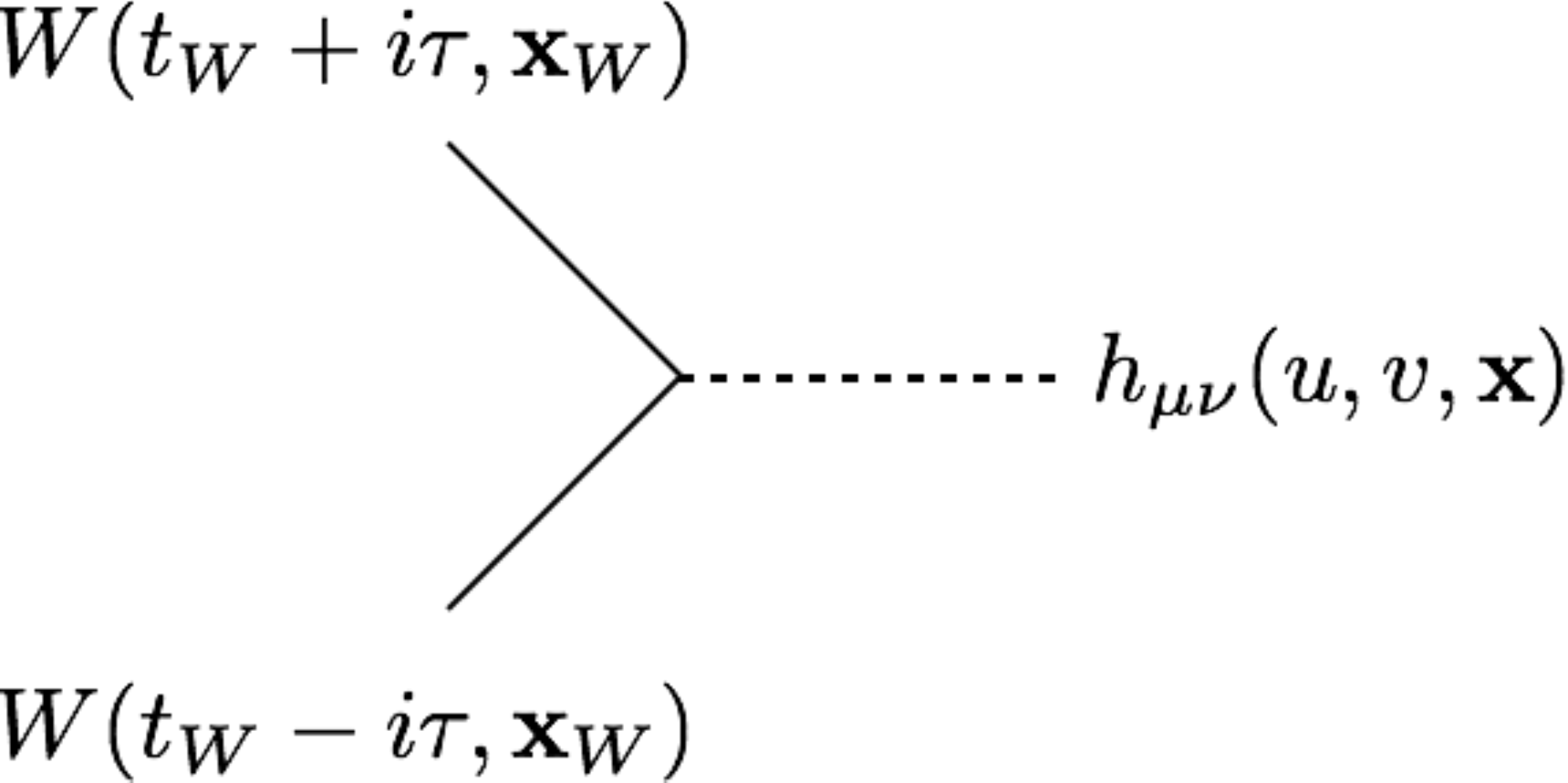}}
 \caption{Three-point diagram as a part of the bulk tree-level graviton exchange diagram. The two straight lines represent the bulk-boundary scalar propagators, and a dotted line represents the bulk-bulk graviton propagator. The interaction region is localized around the horizon because of (\ref{evemt}).}\label{tpd}
\end{figure}

The Lyapunov exponent and butterfly velocity are defined by the exponential behavior of sub-leading term in OTOC which corresponds to the bulk tree-level graviton exchange diagram. To understand the exponential behavior of the tree-level diagram, let us focus on a three-point diagram as shown in Fig.~\ref{tpd}. This three-point diagram corresponds to a bulk three-point function  at tree-level with two $W$ and metric perturbation $h_{\mu\nu}$.  Since $|\psi\rangle$ includes $W$, the three-point function can be expressed as a classical expectation value $\langle\psi|h_{\mu\nu}(u, v, \mathbf{x})|\psi\rangle$. Thus, one can compute the exponential behavior by using classical analysis of the shock wave geometry which is a solution of Einstein equations. 

The localized energy-momentum tensor (\ref{evemt}) changes the initial metric (\ref{tsbhm}) to the shock wave geometry \cite{Aichelburg:1970dh, Dray:1984ha, Sfetsos:1994xa}
\begin{align}
\dd s^2&=2A(uv)\dd u\dd v+B(uv)\dd \mathbf{x}^2-2A(uv)h_g^M(\mathbf{x})e^{\frac{2\pi}{\beta} t_W}\delta (u)\dd u^2\,,
\end{align}
and the dynamics of $h_g^M(\mathbf{x})$ is determined by the Einstein equations with (\ref{evemt}) as follows:
\begin{align}
\left(\square_M-\frac d2A(0)^{-1}B'(0)\right)h_g^M(\mathbf{x})=\frac{B(0)}{2A(0)}P\delta (\mathbf{x}-\mathbf{x}_W)\,, \label{ehg}
\end{align}
where $\square_M$ is the Laplacian on $M$. An isotropic solution of (\ref{ehg}) with $M=\mathbb{R}^d$ is given by \cite{Roberts:2014isa, Blake:2016wvh, Roberts:2016wdl}
\begin{align}
h_g^{\mathbb{R}}(\mathbf{x})\propto \frac{e^{- \mu_g^{\mathbb{R}}|\mathbf{x}-\mathbf{x}_W|}}{|\mathbf{x}-\mathbf{x}_W|^{(d-1)/2}}\,, \;\;\;\; (\mu_g^{\mathbb{R}})^2=\frac{d B'(0)}{2A(0)}\label{shg}\,,
\end{align}
at large distance $|\mathbf{x}-\mathbf{x}_W|\gg 1/\mu_g^{\mathbb{R}}$.

The classical field $h_g^{\mathbb{R}}(\mathbf{x})e^{\frac{2\pi}{\beta} t_W}$ is related to $\langle\psi|h_{uu}|\psi\rangle$ and the bulk tree-level graviton exchange diagram as explained above. Assuming no light higher spin fields $(\ell>2)$ for holography \cite{Heemskerk:2009pn}, graviton exchange is the dominant contribution to the sub-leading term in OTOC. Since the Lyapunov exponent and butterfly velocity can be extracted from the sub-leading term,  we obtain $\lambda_L$ and $v_B$ in the holographic model with $M=\mathbb{R}^d$ from the exponential behavior in $h_g^{\mathbb{R}}(\mathbf{x})e^{\frac{2\pi}{\beta} t_W}$ as follows:
\begin{align}
\lambda_L=\frac{2\pi}{\beta}\,, \;\;\;\;v_B=\frac{2\pi}{\beta  \mu_g^{\mathbb{R}}}\,.
\end{align}

As a relation between quantum chaos and energy dynamics, it was found that $\lambda_L$ and $v_B$ in the holographic model are related to a pole-skipping point of energy density derived from the near-horizon analysis \cite{Blake:2018leo}. Specifically, a component of Einstein equations with $\omega=i\lambda_L$ at the horizon has the same form as the left-hand side of (\ref{ehg}). In the subsequent sections, we will show that similar phenomena occur in scalar and vector exchanges.

We can do the same job on the hyperbolic boundary space $M=\mathbb{H}^d$. 
We set a metric on $\mathbb{H}^d$ as 
\begin{align}
\dd \mathbf{x}^2=\frac{a^2}{\rho^2}\left(\dd \rho^2+\dd \mathbf{x}^2_\perp\right)\,,\label{hmetric}
\end{align}
where $a$ is a length scale of $\mathbb{H}^d$, and $\mathbf{x}_\perp$ are transverse coordinates on $\mathbb{R}^{d-1}$. The Laplacian on (\ref{hmetric}) is given by
\begin{align}
\square_{\mathbb{H}^{d}}=\frac{1}{a^2}\left(\rho^2\partial^2_\rho-(d-2)\rho\partial_\rho+\rho^2\square_{\mathbb{R}^{d-1}}\right)\,,
\end{align}
and the geodesic distance $\mathbf{d}(\mathbf{x}_1, \mathbf{x}_2)$ on (\ref{hmetric}) is defined by
\begin{align}
\cosh \mathbf{d}(\mathbf{x}_1, \mathbf{x}_2):=\frac{\rho_1^2+\rho^2_2+\mathbf{x}_{\perp 12}^2}{2\rho_1\rho_2}\,.\label{Hgeod}
\end{align}
In this paper, we use the metric (\ref{hmetric}) with $a=1$ as in \cite{Haehl:2019eae,Ahn:2019rnq, Ahn:2020bks}  for $M=\mathbb{H}^d$.
Then the solution of \eqref{ehg} with $M=\mathbb{H}^d$ is given by
\begin{align}
h_g^{\mathbb{H}}(\mathbf{x})\propto e^{- \mu_g^{\mathbb{H}}d( \mathbf{x},\mathbf{x}_W)}\,, \;\;\;\; \mu_g^{\mathbb{H}}(\mu_g^{\mathbb{H}}-d+1)=\frac{d B'(0)}{2A(0)}\label{shgh}\,,
\end{align} at large distance $\mathbf{d}(\mathbf{x}_1, \mathbf{x}_2)\gg 1/\mu_g^{\mathbb{H}}$. Thus, the Lyapunov exponent $\lambda_L$ and the butterfly velocity $v_B$ in the holographic model with $M=\mathbb{H}^d$ are
\begin{align}
\lambda_L=\frac{2\pi}{\beta}\,, \;\;\;\;v_B=\frac{2\pi}{\beta  \mu_g^{\mathbb{H}}}\,.
\end{align}
Note that the geometry of the boundary space $M$ only affects the butterfly velocity $v_B$, not the Lyapunov exponent $\lambda_L$.

\section{Scalar exchange}\label{sec3}
In this section, we analyze the exponential behavior of scalar exchange terms in the four-point OTOC by using a simple holographic model. Our computation is a generalization of the method reviewed in Section \ref{sec2}. Specifically, we investigate the exponential behaviors on planar and hyperbolic black holes. We show that the exponential behavior is the same as the one for the leading pole-skipping point derived from the near-horizon analysis of the scalar field.  

\subsection{Exponential behavior with scalar exchange}\label{subsecebs}

On the background geometry (\ref{tsbhm}), we consider actions of the scalar fields
\begin{align}
S&=S_W+S_V+S_\varphi+S_{\textrm{int}W}+S_{\textrm{int}V}\,,\\
S_\varphi&=-\frac{1}{2}\int d^{d+2}x \sqrt{-g} \left(g^{\mu \nu} \partial_{\mu} \varphi \partial_{\nu} \varphi+m_\varphi^2 \varphi^2 \right)\,,\label{svarphi}\\
S_{\textrm{int}W}&=\int d^{d+2}x \sqrt{-g} \left(\lambda_W\phi_W^2\varphi\right)\,,\label{siW}\\
S_{\textrm{int}V}&=\int d^{d+2}x \sqrt{-g} \left(\lambda_V\phi_V^2\varphi\right)\,,\label{siV}
\end{align}
where $S_W$ and $S_V$ are given by (\ref{SW}) and (\ref{SV}).
The bulk three-point interactions (\ref{siW}) and (\ref{siV}) demonstrate that $W$ and $V$ interact with a boundary scalar operator, which corresponds to $\varphi$.
From now on, we evaluate a contribution from the scalar exchange with (\ref{siW}) instead of considering the shock wave geometry.

Similar to the localized energy-momentum tensor (\ref{evemt}) on the  horizon $u=0$ at the late times $t_W\gg\beta$ for graviton exchange, we consider the localized expectation value for the scalar exchange as:
\begin{align}
\frac{\langle\psi|\frac{1}{\sqrt{-g}}\frac{\delta S_{\textrm{int}W}}{\delta \varphi(u, v, \mathbf{x})}|\psi\rangle}{\langle\psi|\psi\rangle}=N_\varphi(t_W)\delta (u)\delta (\mathbf{x}-\mathbf{x}_W)\,.\label{ss}
\end{align}
We want to determine $t_W$-dependence of $N_\varphi(t_W)$ based on \cite{Roberts:2014ifa}. By integrating (\ref{ss}) over $v=0$, we obtain
\begin{align}
N_\varphi(t_W)=\frac{\langle\psi|\int d\mathbf{x} du\frac{1}{\sqrt{-g}}\frac{\delta S_{\textrm{int}W}}{\delta \varphi(u, v, \mathbf{x})}|\psi\rangle|_{v=0}}{\langle\psi|\psi\rangle}\,.\label{ntw}
\end{align}

Let us first start with the denominator. 
The norm $\langle\psi|\psi\rangle$ is expressed as a Klein-Gordon inner product \cite{Shenker:2014cwa, Roberts:2014ifa}
\begin{align}
\langle\psi|\psi\rangle=2iB(0)^{d/2}\int d\mathbf{x} du K(t_W+i\tau, \mathbf{x}_W; u, v, \mathbf{x})^*\partial_uK(t_W+i\tau, \mathbf{x}_W; u, v, \mathbf{x})|_{v=0}\,,
\end{align}
where $K(t_W, \mathbf{x}_W; u, v, \mathbf{x})$ is a bulk-to-boundary propagator of $\phi_W$ which is determined from (\ref{SW}) on (\ref{tsbhm}). Since the black hole metric (\ref{osbhm}) does not depend on $t$, we assume that the propagator $K(t_W, \mathbf{x}_W; u, v, \mathbf{x})$ has a time translation symmetry.
This assumption means that  $K(t_W, \mathbf{x}_W; u, v, \mathbf{x})$ is a function of $t_W-t$, and we can express 
\begin{align}
K(t_W, \mathbf{x}_W; u, v, \mathbf{x})=K(e^{\frac{2\pi}{\beta} t_W}u, e^{-\frac{2\pi}{\beta}t_W}v, \mathbf{x}_W, \mathbf{x})\,,\label{tts}
\end{align}
where we use $u\propto e^{-\frac{2\pi}{\beta}t}$ and $v\propto e^{+\frac{2\pi}{\beta}t}$ derived from (\ref{Kc}). With the time translation symmetry assumption and transformation $u'=e^{\frac{2\pi}{\beta}t_W}u$, we obtain
\begin{align}
\langle\psi|\psi\rangle&=2iB(0)^{d/2}\int d\mathbf{x} duK(e^{\frac{2\pi}{\beta}(t_W+i\tau)}u, 0, \mathbf{x}_W, \mathbf{x})^*\partial_uK(e^{\frac{2\pi}{\beta}(t_W+i\tau)}u, 0, \mathbf{x}_W, \mathbf{x})\notag\\
&=2iB(0)^{d/2}\int d\mathbf{x} du'K(e^{\frac{2\pi}{\beta}i\tau}u', 0, \mathbf{x}_W, \mathbf{x})^*\partial_{u'}K(e^{\frac{2\pi}{\beta} i\tau}u', 0, \mathbf{x}_W, \mathbf{x})\,,\label{psinorm}
\end{align} 
and therefore $\langle\psi|\psi\rangle$ does not depend on $t_W$.

Next, for the numerator, we can estimate that
\begin{align}
\langle\psi|\int d\mathbf{x} du\frac{1}{\sqrt{-g}}\frac{\delta S_{\textrm{int}W}}{\delta \varphi(u, v, \mathbf{x})}|\psi\rangle|_{v=0}=\lambda_W\langle\psi|\int d\mathbf{x} du \phi_W^2(u, 0, \mathbf{x})|\psi\rangle\,.
\end{align}
By contracting $\phi_W$ with the boundary operator $W$ in $|\psi\rangle$, we obtain
\begin{align}
 &\langle\psi|\int d\mathbf{x} du\frac{1}{\sqrt{-g}}\frac{\delta S_{\textrm{int}W}}{\delta \varphi(u, v, \mathbf{x})}|\psi\rangle|_{v=0}\notag\\
 =&2\lambda_W\int d\mathbf{x} duK(e^{\frac{2\pi}{\beta}(t_W+i\tau)}u, 0, \mathbf{x}_W, \mathbf{x})^*K(e^{\frac{2\pi}{\beta}(t_W+i\tau)}u, 0, \mathbf{x}_W, \mathbf{x})\notag\\
 =&2\lambda_We^{-\frac{2\pi}{\beta}t_W}\int d\mathbf{x} du'K(e^{\frac{2\pi}{\beta}i\tau}u', 0, \mathbf{x}_W, \mathbf{x})^*K(e^{\frac{2\pi}{\beta}i\tau}u', 0, \mathbf{x}_W, \mathbf{x})\,,\label{ese}
\end{align}
which is proportional to $e^{-\frac{2\pi}{\beta}t_W}$. From the $t_W$-dependence of (\ref{psinorm}) and (\ref{ese}), it follows that $N_\varphi(t_W)$ in (\ref{ntw}) is proportional to $e^{-\frac{2\pi}{\beta}t_W}$ at the late times: 
\begin{align}
N_\varphi(t_W) \sim  e^{-\frac{2\pi}{\beta}t_W}\,,\label{expbehav}
\end{align}
where we ignore an insignificant coefficient.
Unlike the exponential boost $e^{+\frac{2\pi}{\beta}t_W}$ in graviton exchange, the exponential behavior $e^{-\frac{2\pi}{\beta}t_W}$ in scalar exchange decays at late times $t_W\gg\beta$. This is the reason why scalar exchange is excluded in computations of the Lyapunov exponent with the eikonal approximation.

With (\ref{svarphi})  and (\ref{ss}), an e.o.m.~ of $\varphi$ is
\begin{align}
\frac{1}{\sqrt{-g}}\partial_\mu(\sqrt{-g}g^{\mu\nu}\partial_\nu\varphi)-m_\varphi^2\varphi=-N_\varphi e^{-\frac{2\pi}{\beta} t_W}\delta (u)\delta (\mathbf{x}-\mathbf{x}_W)\,.\label{eomvarphi}
\end{align}
To solve it, we use an ansatz
\begin{align}
\varphi(u, v, \mathbf{x})=e^{-\frac{2\pi}{\beta} t_W}f_\varphi(uv)\delta(u)h_\varphi^M(\mathbf{x})\,.
\end{align}
Without loss of generality, we set $f_\varphi(0)=1$.
After we put this ansatz into \eqref{eomvarphi}, the e.o.m.~ becomes 
\begin{align}
& \left[A^{-1}B^{-d/2}\left(\partial_u(B^{d/2}\partial_v f_\varphi)+\partial_v(B^{d/2}\partial_u f_\varphi)\right)+B^{-1}f_\varphi\square_M \notag-m_\varphi^2f_\varphi\right] \delta (u) h_\varphi^M(\mathbf x) \\
+&A^{-1}B^{-d/2}\left[B^{d/2}(\partial_v f_\varphi) +\partial_v\left(B^{d/2} f_\varphi\right)  \right]  \delta'(u) h_\varphi^M(\mathbf x)=-N_\varphi \delta(u) \delta(\mathbf x-\mathbf{x}_W)\,. 
\end{align}
Using $u\delta'(u)=-\delta (u)$ and $u\delta(u)=0$ \cite{Roberts:2014isa}, we obtain a relation on the horizon $u=0$ 
\begin{align}
&\left[\square_M -m_\varphi^2B(0)-\frac d2A(0)^{-1}B'(0)\right]h_\varphi^M(\mathbf x)=-N_\varphi B(0)\delta(\mathbf x-\mathbf{x}_W)\,, \label{ehs2}
\end{align}
where $ B'(uv):=\partial_{uv}B(uv)$. Note that (\ref{ehs2}) does not depend on derivatives of $f_\varphi(uv)$.
Furthermore, using expressions on the horizon $u=0\,(r=r_0)$:
\begin{align}
    A(uv)\vert_{u\rightarrow0}&=\left.\frac{2}{U'(r_0)^2}\frac{U(r(uv))}{uv}\right\vert_{u\rightarrow0},\\
    B'(uv)\vert_{u\rightarrow0}&=\frac {V'(r_0)}{U'(r_0)}\left.\frac{U(r(uv))}{uv}\right\vert_{u\rightarrow0}\,, 
\end{align}
we can express (\ref{ehs2}) as follows:
\begin{align}
\left(\square_M-m_\varphi^2V(r_0)-d\pi V'(r_0)/\beta\right)h_\varphi^M(\mathbf{x})=-N_\varphi V(r_0)\delta (\mathbf{x}-\mathbf{x}_W)\,.\label{ehs}
\end{align}
This equation determines the exponential behavior in $h_\varphi^M(\mathbf{x})$.

As an explicit example,  an isotropic solution of \eqref{ehs} with $M=\mathbb{R}^d$ is
\begin{align}
h_\varphi^{\mathbb{R}}(\mathbf{x})\propto \frac{e^{- \mu_\varphi^{\mathbb{R}}|\mathbf{x}-\mathbf{x}_W|}}{|\mathbf{x}-\mathbf{x}_W|^{(d-1)/2}}\,,\qquad \left(\mu_\varphi^{\mathbb{R}}\right)^2=m_\varphi^2V(r_0)+\frac{d\pi}{\beta} V'(r_0)\label{srs}
\end{align}
at large distance $|\mathbf{x}-\mathbf{x}_W|\gg 1/\mu_\varphi^{\mathbb{R}}$. In case of $M=\mathbb{H}^d$, a solution of \eqref{ehs} with $SO(d-1,1)$ symmetry is
\begin{align}
h_\varphi^{\mathbb{H}}(\mathbf{x})\propto e^{- \mu_\varphi^{\mathbb{H}} d(\textbf x,\textbf x_W)}\,,\qquad \mu^{\mathbb{H}}_\varphi(\mu^{\mathbb{H}}_\varphi -d+1)=m_\varphi^2V(r_0)+\frac{d\pi}{\beta} V'(r_0)\label{shs}
\end{align}
at large distance $d(\textbf x,\textbf x_W)\gg 1$, where $d(\textbf x,\textbf x_W)$ is the $SO(d-1,1)$ invariant geodesic distance between $\textbf x$ and $\textbf x_W$ in $\mathbb{H}^d$ \eqref{Hgeod}. 

\subsection{Pole-skipping points of the scalar field}\label{subsecnhas}
An alternative method to obtain the Lyapunov exponent $\lambda_L$ and butterfly velocity $v_B$ is to seek the pole-skipping points of energy density. Pole-skipping points are the points in the momentum space that render the Green’s function non-unique: $0/0$. One of the methods to diagnose the pole-skipping points is the near-horizon analysis. This analysis detects the non-uniqueness of the Green's function by the enhancement of the number of free parameters at the horizon $r=r_0$. 
In this section, we review the near-horizon analysis of the minimally-coupled scalar fields \cite{Blake:2019otz} on the general boundary space $M$. 
The near-horizon analysis in the planar space ($M=\mathbb{R}^d$) and the hyperbolic space ($M=\mathbb{H}^d$) has been studied extensively, as we can observe from  \cite{Grozdanov:2019uhi, Ahn:2020bks, Natsuume:2019xcy}.

To perform the near-horizon analysis of the minimally coupled scalar field $\varphi$, we only consider the action \eqref{svarphi}. Its e.o.m.~ is
\begin{align}
   \square\varphi-m_\varphi^2\varphi=0\,,
\end{align}
where $\square$ is the Laplacian corresponding to the general metric.
Using the incoming Eddington-Finkelstein coordinates $v_{EF}=t+r_*$  with the tortoise coordinate defined in \eqref{Kc}, the metric in our purpose is
\begin{equation}
	\dd s^2=-U(r) \dd v_{EF}^2 +2 \dd v_{EF} \dd r+V(r)\dd \mathbf{x}^2\,. \label{Eddmetric}
\end{equation}
Using this metric \eqref{Eddmetric} and the scalar field perturbation of the form $\varphi(v_{EF},r,\mathbf{x})\sim\phi(r,\mathbf{x})e^{-i\omega v_{EF}}$, the e.o.m.~ becomes
\begin{equation}
    \phi''+(UV)^{-1}\left(U'V+d UV'/2-2i\omega V\right) \phi'+(UV)^{-1}\left(\square_M-m_\varphi^2V-i\omega d V'/2\right)\phi=0\,,\label{eom}
\end{equation} 
where $\square_M$ is the Laplacian of the general boundary space $M$, prime is the derivative with respect to $r$, and the arguments of the scalar field $\phi(r,\mathbf x)$ and $U(r),V(r)$ are omitted.

The e.o.m.~\eqref{eom} has a regular singular point at the horizon because 
$U \sim (r-r_0)$ and $V \sim (r-r_0)^0$.  
The general solutions of a second-order differential equation with  regular singular points are well-known, and we seek  the conditions where the solution $\phi$ contains two independent regular solutions. In this case, the solution is determined by two parameters so that the holographic Green's function becomes a function of the ratio of two parameters, yielding a non-unique Green's function. 

As a first step, we examine the Frobenius series solution near the horizon $r=r_0$,
\begin{align}
    \phi(r;\phi_0)=\sum_{n=0} \phi_{n} (r-r_0)^{\alpha +n}\,\quad (\phi_0\neq0)\,.\label{Frob}
\end{align}
Argument $\phi_0$ in $\phi(r;\phi_0)$ denotes the free coefficient of the series which determines all the  other coefficients for a given $\alpha$. 
 After we insert \eqref{Frob} into \eqref{eom}, we can find the so-called indicial equation at the lowest power of $(r-r_0)$, which determines the value of $\alpha$.  
Solving this indicial equation gives two possible values of $\alpha$
\begin{equation}
	\alpha_1=0\,,\quad \alpha_2=i\tilde\omega\,,\label{indicialsol}
\end{equation}
where $i\tilde\omega=\frac{i\omega\beta}{2\pi }=\frac{2i\omega}{U'(r_0)} $.  

The forms of two independent solutions of \eqref{eom} depend on the difference between the two roots of the indicial equation: $\alpha_2-\alpha_1=i\tilde\omega$. The solutions of \eqref{eom} are classified as three cases: $i\tilde\omega$ is i) non-integer ii) zero iii) non-zero integer.
\paragraph{i) $i\tilde\omega$ is non-integer} 
\begin{align*}
    \phi^{(1)}(r;\phi^{(1)}_0)&=\sum_{n=0}^\infty \phi^{(1)}_{n} (r-r_0)^{n}\,,\\
    \phi^{(2)}(r;\phi^{(2)}_0)&=\sum_{n=0}^\infty \phi^{(2)}_{n} (r-r_0)^{i\tilde\omega+n}\,.
\end{align*}
There are two Frobenius series solutions for $\alpha_1=0$ and $\alpha_2=i\tilde\omega$.  As the second solution $\phi^{(2)}$ has non-integer exponents $(r-r_0)^{i\tilde\omega +n}$, $\phi^{(2)}$ is not regular. Thus, the regularity condition picks up only one solution $\phi^{(1)}$. As the solution can be uniquely determined by the single free coefficient $\phi^{(1)}_0$,\footnote{Indeed, this parameter can be set to be $1$, because the equation is linear.} so does the holographic Green's function.
\paragraph{ii) $i\tilde\omega$ is zero} 
\begin{align*}
    &\phi^{(1)}(r;\phi^{(1)}_0)=\sum_{n=1}^\infty \phi^{(1)}_{n} (r-r_0)^{n}+\phi^{(2)}(r;\phi^{(1)}_0)\log(r-r_0)\,,\\
    &\phi^{(2)}(r;\phi^{(2)}_0)=\sum_{n=0}^\infty \phi^{(2)}_{n} (r-r_0)^{n}\,.
\end{align*}
As the first solution $\phi^{(1)}$ contains the $\log$ term, it can be inferred that it is not regular. According to the regularity condition, only the second solution $\phi^{(2)}$ is  allowed. Thus, the holographic Green's function can be uniquely determined. 
\paragraph{iii) $i\tilde\omega$ is  non-zero integer} 
\begin{align}
    &\phi^{(1)}(r;\phi^{(1)}_0)=\sum_{n=0}^\infty \phi^{(1)}_{n} (r-r_0)^{n}+F(\omega,k_i)\phi^{(2)}(r;\phi^{(1)}_0)\log(r-r_0)\,,\label{1stsol}\\
    &\phi^{(2)}(r;\phi^{(2)}_0)=\sum_{n=0}^\infty \phi^{(2)}_{n} (r-r_0)^{i\tilde\omega +n}\,.\label{2ndsol}
\end{align}
As in case ii), there is one solution with the logarithm for $\alpha_1=0$ but the difference is that the factor $F(\omega,k_i)$\footnote{$k_i$ are related to the eigenvalues of $\square_M$ as $\square_M\phi=\lambda(k_i)\phi$. $\lambda(k_i)$ is some function that depends on the boundary space $M$ and $\lambda(k_i)=-\sum_i k_i^2$ at $M=\mathbb R^d$ for example. In this section, we use $\square_M$ and $\lambda(k_i)$ interchangeably. \label{kvalue}} is multiplied to the logarithm. The regularity conditions for each solution are: a) the first solution \eqref{1stsol} is regular only if $F(\omega,k_i)$ is zero, b) the second solution \eqref{2ndsol} is regular only if $i\tilde\omega$ is a positive integer. It means that when $i\tilde\omega$ is a positive integer and $F(\omega,k_i)=0$, there are two free parameters $\phi_0^{(1)},\phi_0^{(2)}$.
Thus, in this case, the holographic Green's function is not uniquely defined and such points are called pole-skipping points.

The leading (smallest) pole-skipping point is $i\tilde\omega=1$ ($\omega=-i2\pi /\beta$) and 
\begin{equation}
    F(-i2\pi /\beta,k_i) \sim \square_M-m_\varphi^2V(r_0)-\pi d V'(r_0)/\beta \,,
\end{equation}
which can be obtained by plugging the form of \eqref{1stsol} into the e.o.m.~\eqref{eom}. Here, $k_i$ is encoded in $\square_M$ as explained in footnote \ref{kvalue}. 

In summary, the pole-sipping conditions are
\begin{equation}
    \omega=-i2\pi /\beta,\qquad\square_M-m_\varphi^2V(r_0)-\pi d V'(r_0)/\beta=0\,.\label{logNHA}
\end{equation}
 These conditions coincide with the coefficient of the exponential behavior of the scalar field $e^{-\frac{2\pi }{\beta}t_W}$ \eqref{expbehav} (the massive scalar field in near-horizon analysis behaves like $\varphi\sim e^{-i\omega v_{EF}}=e^{-\frac{2\pi}{\beta}v_{EF}}$) and the condition for the spatial part $h^M_\varphi(\mathbf x)$ \eqref{ehs} obtained by the scalar exchange in the previous subsection.

An alternative way to obtain the leading pole-skipping point is as follows. 
By plugging \eqref{Frob} to the e.o.m.~ at the near-horizon limit,  we can get the expression at the lowest order as
\begin{align}
    \left(\square_M-m_\varphi^2V(r_0)-i\omega d V'(r_0)/2\right)\phi_0+(4\pi /\beta-2i\omega)V(r_0)\phi_1=0\label{NHAcoeff}\,.
\end{align}
One can observe that $\phi_0$ and $\phi_1$ cannot be determined when $\omega=-i2\pi /\beta$ and $\square_M-m_\varphi^2V(r_0)-i\omega d V'(r_0)/2=0$.
These conditions give the leading pole-skipping points of the massive scalar field again.

We leave some comments on the similarities and differences between the two analysis methods in Subsections \ref{subsecebs} and \ref{subsecnhas}.
\begin{itemize}
\item Both analysis methods  depend strongly on the metric at the black hole horizon.
\item The procedure to determine the late time behavior $e^{-\frac{2\pi}{\beta}t_W}$ in Subsection \ref{subsecebs} depends on the three-point interaction (\ref{siW}) based on (\ref{ese}). On the other hand, in Subsection \ref{subsecnhas}, the late time behavior can be determined from the e.o.m.~ of $\varphi$ only. 
\item In the four-point OTOC with $M=\mathbb{R}^d$, space propagation is expressed in terms of  isotropic propagation $e^{ik|\mathbf{x}|}$ as in (\ref{shs}). On the other hand, in the near-horizon analysis, we often use Fourier expansion with $e^{ik_ix^i}$ instead of $e^{ik|\mathbf{x}|}$. See, for instance, \cite{Blake:2018leo}.
\end{itemize}

\section{Vector exchange}\label{sec4}

Here, we study the exponential behaviors of the vector exchange terms by considering the interactions of complex scalar fields with a vector field.  As in the case of scalar fields, we demonstrate that they are related to the leading pole-skipping points in the near-horizon analysis of the vector field. 

\subsection{Exponential behavior with vector exchange}

On the black hole background metric (\ref{tsbhm}), we consider actions of the complex scalar fields and a vector field
\begin{align}
S&=S_W+S_V+S_A+S_{\textrm{int}W}+S_{\textrm{int}V},\\
S_W&=-\int d^{d+2}x \sqrt{-g} \left(g^{\mu \nu} \partial_{\mu} \phi_W^\dagger \partial_{\nu} \phi_W+m_W^2 |\phi_W|^2 \right)\,,\label{SaW}\\
S_V&=-\int d^{d+2}x \sqrt{-g} \left(g^{\mu \nu} \partial_{\mu} \phi_V^\dagger \partial_{\nu} \phi_V+m_V^2 |\phi_V|^2 \right)\,,\\
S_A&=-\int d^{d+2}x \sqrt{-g} \left(\frac{1}{4}F_{\mu\nu}F^{\mu\nu}+\frac{1}{2}m_A^2 A_\mu A^\mu \right)\,,\label{sa}\\
S_{\textrm{int}W}&=iq_W\int d^{d+2}x \sqrt{-g} A^\mu\left[(\partial_\mu\phi_W)\phi_W^\dagger-\phi_W\partial_\mu\phi_W^\dagger\right],\label{saiW}\\
S_{\textrm{int}V}&=iq_V\int d^{d+2}x \sqrt{-g} A^\mu\left[(\partial_\mu\phi_V)\phi_V^\dagger-\phi_V\partial_\mu\phi_V^\dagger\right].\label{saiV}
\end{align}
The bulk complex scalar fields $\phi_W$ and $\phi_V$ are dual to the boundary operators $W$ and $V$ in the four-point OTOC $\langle W^\dagger(t_W, \mathbf{x}_W)V^\dagger(0,\mathbf{x}_V)W(t_W, \mathbf{x}_W)V(0,\mathbf{x}_V)\rangle$. The bulk  interactions (\ref{saiW}) and (\ref{saiV}) as three-point interaction\footnote{We do not consider four-point interaction such as $A_\mu A^\mu|\phi_W|^2$ because it is not relevant to bulk tree level diagrams for $\langle W^\dagger(t_W, \mathbf{x}_W)V^\dagger(0,\mathbf{x}_V)W(t_W, \mathbf{x}_W)V(0,\mathbf{x}_V)\rangle$.} mean that $W$ and $V$ interact with a boundary vector operator, which is dual to $A_\mu$. 

As (\ref{ss}) in the previous section, we consider a localized expectation value at late times $t_W\gg \beta$ for vector exchange
\begin{align}
\frac{\langle\psi|\frac{1}{\sqrt{-g}}\frac{\delta S_{\textrm{int}W}}{\delta A^\mu(u, v, \mathbf{x})}|\psi\rangle}{\langle\psi|\psi\rangle}=N_{A^{\mu}}(t_W)\delta (u)\delta (\mathbf{x}-\mathbf{x}_W),\label{vsuc}
\end{align}
and determine $t_W$-dependence of $N_{A^{\mu}}(t_W)$. With (\ref{saiW}), for $A^u$, we obtain
\begin{align}
&\langle\psi|\int d\mathbf{x} du\frac{1}{\sqrt{-g}}\frac{\delta S_{\textrm{int}W}}{\delta A^u(u, v, \mathbf{x})}|\psi\rangle|_{v=0}=iq_W\langle\psi|\int d\mathbf{x} du \left[(\partial_u\phi_W)\phi_W^\dagger-\phi_W\partial_u\phi_W^\dagger\right]|\psi\rangle|_{v=0}\notag\\
=&2iq_W\int d\mathbf{x} du \left(\partial_uK(e^{\frac{2\pi}{\beta}(t_W+i\tau)}u, 0, \mathbf{x}_W, \mathbf{x})^*\right)K(e^{\frac{2\pi}{\beta}(t_W+i\tau)}u, 0, \mathbf{x}_W, \mathbf{x})\notag\\
=&2iq_W\int d\mathbf{x} du' \left(\partial_{u'}K(e^{\frac{2\pi}{\beta}i\tau}u', 0, \mathbf{x}_W, \mathbf{x})^*\right)K(e^{\frac{2\pi}{\beta}i\tau}u', 0, \mathbf{x}_W, \mathbf{x}),\label{eve}
\end{align}
where we use the time translation symmetry (\ref{tts}) and $u'=e^{\frac{2\pi}{\beta}t_W}u$. Because of $\partial_u$, (\ref{eve}) does not depend on $t_W$ unlike (\ref{ese}).
Since (\ref{eve}) and $\langle\psi|\psi\rangle$ do not depend on $t_W$, we conclude that $N_{A^u}(t_W)$ in (\ref{vsuc}) for vector exchange does not depend on $t_W$ at the late times: 
\begin{equation}
    N_{A^u}(t_W) \sim e^{0\cdot t_W}  \sim \mathcal{O}(1).
\end{equation}
Similarly, one can estimate $t_W$-dependence of the other components
\begin{align}
N_{A^\mu} \sim \mathcal{O}(e^{-\frac{2\pi}{\beta}t_W}) \qquad(\mu\ne u).\label{vsoc}
\end{align}
This difference of the $t_W$-dependence between components seems to be related to different values  of $\omega$ at the leading pole-skipping points in different channels.

With (\ref{sa}), (\ref{vsuc}) and (\ref{vsoc}), an e.o.m.~ of $A_\nu$ is 
\begin{align}
\nabla^\mu F_{\mu\nu}-m^2_AA_\nu=-\delta_\nu^uN_{A^{\nu}}\delta(u)\delta (\mathbf{x}-\mathbf{x}_W),
\end{align}
where we ignore $\mathcal{O}(e^{-\frac{2\pi}{\beta}t_W})$ terms. To solve the above equation, we use an ansatz
\begin{align}
A_u=&f_A(uv)\delta(u)h_A^M(\mathbf{x}),\notag\\
A_\nu=&\mathcal{O}(e^{-\frac{2\pi}{\beta}t_W})\sim0 \qquad (\nu\ne u),
\end{align}
where we set $f_A(0)=1$. Using $u\delta'(u)=-\delta(u)$ and $u\delta(u)=0$, it turns out that the e.o.m.~ of the vector field has only $\delta(u)$ dependent terms. On the horizon $u=0$, the equation with $\nu=u$ becomes\footnote{The other components become trivial because of $u\delta(u)=0$.} 
\begin{align}
    (\square_M-m_A^2V(r_0)) h_A^M(\mathbf x)=-N_{A^{ u}}V(r_0)\delta(\mathbf{x}-\mathbf{x}_W)\,.\label{ehv}
\end{align}
This relation determines the spatial exponential behavior in $h_A^M(\mathbf x)$.

When $M=\mathbb{R}^d$, an isotropic solution of \eqref{ehv} is
\begin{align}
h_A^{\mathbb{R}}(\mathbf{x})\propto \frac{e^{- \mu_A^{\mathbb R}|\mathbf{x}-\mathbf{x}_W|}}{|\mathbf{x}-\mathbf{x}_W|^{(d-1)/2}}\,,\qquad \left(\mu_A^{\mathbb{R}}\right)^2=m_A^2V(r_0)\label{srv}
\end{align}
at large distance $|\mathbf{x}-\mathbf{x}_W|\gg 1/\mu_A^{\mathbb R}$. For $M=\mathbb{H}^d$, an $SO(d-1, 1)$ invariant  solution of \eqref{ehv} is
\begin{align}
h_A^{\mathbb{H}}(\mathbf{x})\propto e^{- \mu_A^{\mathbb H} d(\textbf x,\textbf x_W)}\,,\qquad \mu_A^{\mathbb{H}}(\mu_A^{\mathbb{H}}-d+1)=m_A^2V(r_0)\label{shv}
\end{align}
at large distance $d(\textbf x,\textbf x_W)\gg 1$.

\subsection{Pole-skipping points of the vector field}\label{subsecnhav}

The action of the bulk vector field \eqref{sa} yields the e.o.m.~
\begin{align}
	\nabla_\mu F^{\mu\nu	}-m_A^2A^\nu=0\,.
\end{align}
With the Eddington-Finkelstein coordinate \eqref{Eddmetric}, the e.o.m.~ of $\nu=v_{EF},r$ components with the Lorenz condition decouple from the other components $\nu\neq v_{EF},r$. Such a sector is termed as a diffusive or longitudinal channel and is relevant to the leading pole-skipping points for the vector field \cite{Blake:2019otz, Natsuume:2019xcy, Grozdanov:2019uhi}. Thus, we only consider the longitudinal channel in this section.

By using an  ansatz $A_\mu(v_{EF},r,\mathbf x)\sim \mathcal A_\mu(r,\mathbf x)e^{-i\omega v_{EF}}$, the e.o.m.~ with $\nu=v_{EF},r$ components are
\begin{align}
&V^{-1}\left(\square_M-m_A^2V-i\omega\frac d2 V'\right)\mathcal A_r-i\omega\mathcal A_r'-\mathcal A_{v_{EF}}''-\frac d2 V' V^{-1}\mathcal A_{v_{EF}}' -(\nabla_i\mathcal A^i)'=0\,,\label{Aveom}\\
&V^{-1}\left(\square_M-m_A^2V\right)\mathcal A_{v_{EF}}+ UV^{-1}\left(\square_M-m_A^2V +\omega^2V  U^{-1}\right)\mathcal A_r-i\omega\mathcal A_{v_{EF}}'\nonumber\\ 
&\qquad\qquad\qquad\qquad\qquad\qquad\qquad\qquad\qquad\qquad\quad+i\omega\nabla_i\mathcal A^i - U(\nabla_i\mathcal A^i)'	=0\,,\label{Areom}
\end{align}
where $i$ is the index of the coordinates $\mathbf x$, and prime is the derivative with respect to $r$. We omitted the arguments of the vector fields $\mathcal A_\mu(r,\mathbf x)$ and $U(r),V(r)$ for  compact expressions. Using the Lorenz condition
\begin{equation}
	\nabla_\mu A^\mu=\left(\nabla_i\mathcal A^i-i\omega\mathcal A_r+\mathcal A_{v_{EF}}'+ ( U\mathcal A_r)'+\frac d2 V^{-1}V'  (\mathcal A_{v_{EF}}+ U\mathcal  A_r)\right) e^{-i\omega v_{EF}}=0\,,
\end{equation}
we can arrange \eqref{Aveom} and \eqref{Areom} only with the variables $\mathcal A_{v_{EF}}(r,\mathbf x)$ and $\mathcal A_r(r,\mathbf x)$. 

Now we use near-horizon analysis to determine the leading pole-skipping points of the vector field. 
First, by inserting the series form of $\mathcal A_\mu$ as
\begin{equation}
	\mathcal A_\mu(r,\mathbf x)=\sum_{n=0}^\infty\mathcal A^n_\mu(\mathbf x)(r-r_0)^{n}\quad (\mathcal A_\mu^0\neq0)\,,
\end{equation}
\eqref{Aveom} and \eqref{Areom} at the lowest order become
\begin{align}
&\frac d2\left(\frac{V''(r_0)}{V(r_0)}-\frac{V'(r_0)^2}{V(r_0)^2}\right)\mathcal A_{v_{EF}}^0\nonumber\\
&\quad+V^{-1}(r_0)\left(\square_M-m_A^2V(r_0)+U''(r_0)V(r_0)+(4\pi/\beta-i\omega)\frac d2V'(r_0)\right)\mathcal A_r^0\nonumber\\
&\quad+2(4\pi/\beta-i\omega)\mathcal A_r^1=0\,,\label{NHAveom}\\
&V^{-1}(r_0)\left(\square_M-m_A^2V(r_0)-i\omega\frac d2 V'(r_0)\right)\mathcal A_{v_{EF}}^0	-\frac{4\pi i\omega}\beta \mathcal A_r^0-2i\omega A_{v_{EF}}^1=0\,.\label{NHAreom}
\end{align}
For most cases, if $\mathcal A_{v_{EF}}^0$ and $\mathcal A_r^0$ are fixed, the higher-order fields $\mathcal A_{v_{EF}}^1, \mathcal A_r^1$ can be determined by \eqref{NHAveom} and \eqref{NHAreom}. However, in case of $\omega=0$ and $\square_M-m_A^2V(r_0)=0$, $\mathcal A_{v_{EF}}^1$ cannot be determined by $\mathcal A_{v_{EF}}^0$ from \eqref{NHAreom}. It is the enhancement of the number of free parameters. This condition determines the leading pole-skipping points of the vector field and coincides with the exponential behavior of the vector exchange: independence of $t_W$ at the late times and \eqref{ehv} in the previous subsection.

\section{Summary and discussion}\label{sec5}

We have studied exponential behaviors of scalar and vector exchange terms in four-point OTOCs by investigating simple holographic models. We have shown that the exponential behaviors are related to special points in the near-horizon e.o.m.~ for scalar and the vector fields. Let us summarize the results of our calculations. 

\paragraph{Scalar field} 
The exponential behavior in scalar exchange at late times $t_W\gg\beta$ is $h^M_\varphi(\mathbf{x})e^{-\frac{2\pi}{\beta}t_W}$, where $h^M_\varphi(\mathbf{x})$ is determined by (\ref{ehs}). In the near-horizon analysis, the leading pole-skipping points are determined by the non-uniqueness conditions of the holographic Green's function in (\ref{logNHA}), and they coincide with $e^{-\frac{2\pi}{\beta}t_W}$ and (\ref{ehs}).

\paragraph{Vector field} 
The exponential behavior in vector exchange at the late times does not depend on $t_W$ as $h^M_A(\mathbf{x})$, where $h^M_A(\mathbf{x})$ is determined by (\ref{ehv}). This time-independence and (\ref{ehv}) can be extracted  from the near-horizon analysis by imposing that (\ref{NHAreom}) is trivial at $\omega=0$.

It is well-known that the four-point correlation functions in CFTs can be expanded in terms of conformal blocks. Although the Lyapunov exponent in holographic CFTs is computed from the conformal block with the energy-momentum tensor exchange \cite{Perlmutter:2016pkf}, the four-point OTOCs have sub-leading contributions from the conformal blocks with the exchange of other fields. Similarly, the four-point OTOCs computed from the holographic method may have sub-leading contributions from the bulk exchange other than graviton. We have computed the exponential behaviors of such sub-leading contributions from scalar and the vector fields. We have also shown that their exponential behaviors are related to the leading pole-skipping points derived from the near-horizon analysis.  

Originally, the pole-skipping phenomena in the near-horizon analysis are relations between the exponential behavior in graviton exchange in OTOC and the near-horizon Einstein's equations \cite{Blake:2018leo}.
Our results imply a generalization  of the pole-skipping phenomena for arbitrary bosonic fields. Note that the pole-skipping phenomena of the graviton or the energy-momentum tensor are related to maximal chaos. However, the pole-skipping phenomena of other fields are not related to maximal chaos as one can see from the $t_W$-dependence of exponential behaviors.

We can now deliberate upon some future studies that can be conducted on this topic. One future direction is to study more complicated holographic actions which are proposed from the viewpoint of AdS/CMT.  For example, 
the dilaton coupling was considered in \cite{Blake:2018leo}. Another future direction is to change the decomposition of the vector field in the near-horizon analysis for comparison with the exponential behavior. It may be useful to decompose the vector field in the near-horizon analysis by Kruskal coordinates. 
It would also be interesting to compute the exponential behavior with the exchange of the other channel in the vector field, although it is a sub-leading term  in vector exchange at the late times.
In the four-point OTOCs with four scalar operators, fermion exchange is forbidden from the fermion number conservation. However, the fermion exchange is possible in the OTOCs with two scalars and two fermions, and we expect pole-skipping points of fermions \cite{Ceplak:2019ymw} to be related with the fermion exchange terms in such OTOCs.

\acknowledgments
We would like to thank Yongjun Ahn, Viktor Jahnke and Chang-Woo Ji for
valuable discussions and comments.
This work was supported by Basic Science Research Program through the National Research Foundation of Korea(NRF) funded by the Ministry of Science, ICT \& Future
Planning (NRF-2017R1A2B4004810) and the GIST Research Institute(GRI) grant funded by the GIST in 2020. 
M. Nishida was supported by Basic Science Research Program through the National Research Foundation of Korea(NRF) funded by the Ministry of Education(NRF-2020R1I1A1A01072726).


\bibliographystyle{JHEP}

\begin{thebibliography}{10}

\bibitem{Shenker:2013pqa}
S.~H. Shenker and D.~Stanford, \emph{{Black holes and the butterfly effect}},
  \href{http://dx.doi.org/10.1007/JHEP03(2014)067}{\emph{JHEP} {\bf 03} (2014)
  067}, [\href{http://arxiv.org/abs/1306.0622}{{\tt 1306.0622}}].

\bibitem{Roberts:2014isa}
D.~A. Roberts, D.~Stanford and L.~Susskind, \emph{{Localized shocks}},
  \href{http://dx.doi.org/10.1007/JHEP03(2015)051}{\emph{JHEP} {\bf 03} (2015)
  051}, [\href{http://arxiv.org/abs/1409.8180}{{\tt 1409.8180}}].

\bibitem{Roberts:2014ifa}
D.~A. Roberts and D.~Stanford, \emph{{Two-dimensional conformal field theory
  and the butterfly effect}},
  \href{http://dx.doi.org/10.1103/PhysRevLett.115.131603}{\emph{Phys. Rev.
  Lett.} {\bf 115} (2015) 131603}, [\href{http://arxiv.org/abs/1412.5123}{{\tt
  1412.5123}}].

\bibitem{Shenker:2014cwa}
S.~H. Shenker and D.~Stanford, \emph{{Stringy effects in scrambling}},
  \href{http://dx.doi.org/10.1007/JHEP05(2015)132}{\emph{JHEP} {\bf 05} (2015)
  132}, [\href{http://arxiv.org/abs/1412.6087}{{\tt 1412.6087}}].

\bibitem{Kitaev-2014}
A.~Kitaev, \emph{{A simple model of quantum holography}}, {\emph{{}} (2015)
  \url{http://online.kitp.ucsb.edu/online/entangled15/kitaev/},
  \url{http://online.kitp.ucsb.edu/online/entangled15/kitaev2/}, Talks at KITP,
  April 7, 2015 and May 27, (2015)}.

\bibitem{Perlmutter:2016pkf}
E.~Perlmutter, \emph{{Bounding the Space of Holographic CFTs with Chaos}},
  \href{http://dx.doi.org/10.1007/JHEP10(2016)069}{\emph{JHEP} {\bf 10} (2016)
  069}, [\href{http://arxiv.org/abs/1602.08272}{{\tt 1602.08272}}].

\bibitem{larkin1969quasiclassical}
A.~Larkin and Y.~N. Ovchinnikov, \emph{Quasiclassical method in the theory of
  superconductivity}, {\emph{Sov Phys JETP} {\bf 28} (1969) 1200--1205}.

\bibitem{Maldacena:2015waa}
J.~Maldacena, S.~H. Shenker and D.~Stanford, \emph{{A bound on chaos}},
  \href{http://dx.doi.org/10.1007/JHEP08(2016)106}{\emph{JHEP} {\bf 08} (2016)
  106}, [\href{http://arxiv.org/abs/1503.01409}{{\tt 1503.01409}}].

\bibitem{Grozdanov:2017ajz}
S.~Grozdanov, K.~Schalm and V.~Scopelliti, \emph{{Black hole scrambling from
  hydrodynamics}},
  \href{http://dx.doi.org/10.1103/PhysRevLett.120.231601}{\emph{Phys. Rev.
  Lett.} {\bf 120} (2018) 231601}, [\href{http://arxiv.org/abs/1710.00921}{{\tt
  1710.00921}}].

\bibitem{Blake:2017ris}
M.~Blake, H.~Lee and H.~Liu, \emph{{A quantum hydrodynamical description for
  scrambling and many-body chaos}},
  \href{http://dx.doi.org/10.1007/JHEP10(2018)127}{\emph{JHEP} {\bf 10} (2018)
  127}, [\href{http://arxiv.org/abs/1801.00010}{{\tt 1801.00010}}].

\bibitem{Choi:2020tdj}
C.~Choi, M.~Mezei and G.~S\'arosi, \emph{{Pole skipping away from maximal
  chaos}},  \href{http://arxiv.org/abs/2010.08558}{{\tt 2010.08558}}.

\bibitem{Blake:2018leo}
M.~Blake, R.~A. Davison, S.~Grozdanov and H.~Liu, \emph{{Many-body chaos and
  energy dynamics in holography}},
  \href{http://dx.doi.org/10.1007/JHEP10(2018)035}{\emph{JHEP} {\bf 10} (2018)
  035}, [\href{http://arxiv.org/abs/1809.01169}{{\tt 1809.01169}}].

\bibitem{Grozdanov:2019uhi}
S.~Grozdanov, P.~K. Kovtun, A.~O. Starinets and P.~Tadi\'{c}, \emph{{The
  complex life of hydrodynamic modes}},
  \href{http://dx.doi.org/10.1007/JHEP11(2019)097}{\emph{JHEP} {\bf 11} (2019)
  097}, [\href{http://arxiv.org/abs/1904.12862}{{\tt 1904.12862}}].

\bibitem{Blake:2019otz}
M.~Blake, R.~A. Davison and D.~Vegh, \emph{{Horizon constraints on holographic
  Green's functions}},
  \href{http://dx.doi.org/10.1007/JHEP01(2020)077}{\emph{JHEP} {\bf 01} (2020)
  077}, [\href{http://arxiv.org/abs/1904.12883}{{\tt 1904.12883}}].

\bibitem{Natsuume:2019xcy}
M.~Natsuume and T.~Okamura, \emph{{Nonuniqueness of Green's functions at
  special points}},
  \href{http://dx.doi.org/10.1007/JHEP12(2019)139}{\emph{JHEP} {\bf 12} (2019)
  139}, [\href{http://arxiv.org/abs/1905.12015}{{\tt 1905.12015}}].

\bibitem{Natsuume:2019vcv}
M.~Natsuume and T.~Okamura, \emph{{Pole-skipping with finite-coupling
  corrections}},
  \href{http://dx.doi.org/10.1103/PhysRevD.100.126012}{\emph{Phys. Rev.} {\bf
  D100} (2019) 126012}, [\href{http://arxiv.org/abs/1909.09168}{{\tt
  1909.09168}}].

\bibitem{Wu:2019esr}
X.~Wu, \emph{{Higher curvature corrections to pole-skipping}},
  \href{http://dx.doi.org/10.1007/JHEP12(2019)140}{\emph{JHEP} {\bf 12} (2019)
  140}, [\href{http://arxiv.org/abs/1909.10223}{{\tt 1909.10223}}].

\bibitem{Ceplak:2019ymw}
N.~Ceplak, K.~Ramdial and D.~Vegh, \emph{{Fermionic pole-skipping in
  holography}},  \href{http://arxiv.org/abs/1910.02975}{{\tt 1910.02975}}.

\bibitem{Abbasi:2019rhy}
N.~Abbasi and J.~Tabatabaei, \emph{{Quantum chaos, pole-skipping and
  hydrodynamics in a holographic system with chiral anomaly}},
  \href{http://dx.doi.org/10.1007/JHEP03(2020)050}{\emph{JHEP} {\bf 03} (2020)
  050}, [\href{http://arxiv.org/abs/1910.13696}{{\tt 1910.13696}}].

\bibitem{Ahn:2020bks}
Y.~Ahn, V.~Jahnke, H.-S. Jeong, K.-Y. Kim, K.-S. Lee and M.~Nishida,
  \emph{{Pole-skipping of scalar and vector fields in hyperbolic space:
  conformal blocks and holography}},
  \href{http://dx.doi.org/10.1007/JHEP09(2020)111}{\emph{JHEP} {\bf 09} (2020)
  111}, [\href{http://arxiv.org/abs/2006.00974}{{\tt 2006.00974}}].

\bibitem{Ahn:2020baf}
Y.~Ahn, V.~Jahnke, H.-S. Jeong, K.-Y. Kim, K.-S. Lee and M.~Nishida,
  \emph{{Classifying pole-skipping points}},
  \href{http://arxiv.org/abs/2010.16166}{{\tt 2010.16166}}.

\bibitem{Natsuume:2020snz}
M.~Natsuume and T.~Okamura, \emph{{Pole-skipping and zero temperature}},
  \href{http://arxiv.org/abs/2011.10093}{{\tt 2011.10093}}.

\bibitem{Grozdanov:2018kkt}
S.~Grozdanov, \emph{{On the connection between hydrodynamics and quantum chaos
  in holographic theories with stringy corrections}},
  \href{http://dx.doi.org/10.1007/JHEP01(2019)048}{\emph{JHEP} {\bf 01} (2019)
  048}, [\href{http://arxiv.org/abs/1811.09641}{{\tt 1811.09641}}].

\bibitem{Li:2019bgc}
W.~Li, S.~Lin and J.~Mei, \emph{{Thermal diffusion and quantum chaos in neutral
  magnetized plasma}},
  \href{http://dx.doi.org/10.1103/PhysRevD.100.046012}{\emph{Phys. Rev. D} {\bf
  100} (2019) 046012}, [\href{http://arxiv.org/abs/1905.07684}{{\tt
  1905.07684}}].

\bibitem{Natsuume:2019sfp}
M.~Natsuume and T.~Okamura, \emph{{Holographic chaos, pole-skipping, and
  regularity}}, \href{http://dx.doi.org/10.1093/ptep/ptz155}{\emph{PTEP} {\bf
  2020} (2020) 013B07}, [\href{http://arxiv.org/abs/1905.12014}{{\tt
  1905.12014}}].

\bibitem{Ahn:2019rnq}
Y.~Ahn, V.~Jahnke, H.-S. Jeong and K.-Y. Kim, \emph{{Scrambling in Hyperbolic
  Black Holes: shock waves and pole-skipping}},
  \href{http://dx.doi.org/10.1007/JHEP10(2019)257}{\emph{JHEP} {\bf 10} (2019)
  257}, [\href{http://arxiv.org/abs/1907.08030}{{\tt 1907.08030}}].

\bibitem{Liu:2020yaf}
Y.~Liu and A.~Raju, \emph{{Quantum Chaos in Topologically Massive Gravity}},
  \href{http://arxiv.org/abs/2005.08508}{{\tt 2005.08508}}.

\bibitem{Abbasi:2020ykq}
N.~Abbasi and S.~Tahery, \emph{{Complexified quasinormal modes and the
  pole-skipping in a holographic system at finite chemical potential}},
  \href{http://arxiv.org/abs/2007.10024}{{\tt 2007.10024}}.

\bibitem{Jansen:2020hfd}
A.~Jansen and C.~Pantelidou, \emph{{Quasinormal modes in charged fluids at
  complex momentum}},  \href{http://arxiv.org/abs/2007.14418}{{\tt
  2007.14418}}.

\bibitem{Grozdanov:2020koi}
S.~Grozdanov, \emph{{Bounds on transport from univalence and pole-skipping}},
  \href{http://arxiv.org/abs/2008.00888}{{\tt 2008.00888}}.

\bibitem{Blake:2016wvh}
M.~Blake, \emph{{Universal Charge Diffusion and the Butterfly Effect in
  Holographic Theories}},
  \href{http://dx.doi.org/10.1103/PhysRevLett.117.091601}{\emph{Phys. Rev.
  Lett.} {\bf 117} (2016) 091601}, [\href{http://arxiv.org/abs/1603.08510}{{\tt
  1603.08510}}].

\bibitem{Roberts:2016wdl}
D.~A. Roberts and B.~Swingle, \emph{{Lieb-Robinson Bound and the Butterfly
  Effect in Quantum Field Theories}},
  \href{http://dx.doi.org/10.1103/PhysRevLett.117.091602}{\emph{Phys. Rev.
  Lett.} {\bf 117} (2016) 091602}, [\href{http://arxiv.org/abs/1603.09298}{{\tt
  1603.09298}}].

\bibitem{Haehl:2019eae}
F.~M. Haehl, W.~Reeves and M.~Rozali, \emph{{Reparametrization modes, shadow
  operators, and quantum chaos in higher-dimensional CFTs}},
  \href{http://dx.doi.org/10.1007/JHEP11(2019)102}{\emph{JHEP} {\bf 11} (2019)
  102}, [\href{http://arxiv.org/abs/1909.05847}{{\tt 1909.05847}}].

\bibitem{Haehl:2018izb}
F.~M. Haehl and M.~Rozali, \emph{{Effective Field Theory for Chaotic CFTs}},
  \href{http://dx.doi.org/10.1007/JHEP10(2018)118}{\emph{JHEP} {\bf 10} (2018)
  118}, [\href{http://arxiv.org/abs/1808.02898}{{\tt 1808.02898}}].

\bibitem{Das:2019tga}
S.~Das, B.~Ezhuthachan and A.~Kundu, \emph{{Real time dynamics from low point
  correlators in 2d BCFT}},
  \href{http://dx.doi.org/10.1007/JHEP12(2019)141}{\emph{JHEP} {\bf 12} (2019)
  141}, [\href{http://arxiv.org/abs/1907.08763}{{\tt 1907.08763}}].

\bibitem{Ramirez:2020qer}
D.~M. Ramirez, \emph{{Chaos and pole skipping in CFT$_2$}},
  \href{http://arxiv.org/abs/2009.00500}{{\tt 2009.00500}}.

\bibitem{Afkhami-Jeddi:2017rmx}
N.~Afkhami-Jeddi, T.~Hartman, S.~Kundu and A.~Tajdini, \emph{{Shockwaves from
  the Operator Product Expansion}},
  \href{http://dx.doi.org/10.1007/JHEP03(2019)201}{\emph{JHEP} {\bf 03} (2019)
  201}, [\href{http://arxiv.org/abs/1709.03597}{{\tt 1709.03597}}].

\bibitem{Aichelburg:1970dh}
P.~Aichelburg and R.~Sexl, \emph{{On the Gravitational field of a massless
  particle}}, \href{http://dx.doi.org/10.1007/BF00758149}{\emph{Gen. Rel.
  Grav.} {\bf 2} (1971) 303--312}.

\bibitem{Dray:1984ha}
T.~Dray and G.~'t~Hooft, \emph{{The Gravitational Shock Wave of a Massless
  Particle}}, \href{http://dx.doi.org/10.1016/0550-3213(85)90525-5}{\emph{Nucl.
  Phys. B} {\bf 253} (1985) 173--188}.

\bibitem{Sfetsos:1994xa}
K.~Sfetsos, \emph{{On gravitational shock waves in curved space-times}},
  \href{http://dx.doi.org/10.1016/0550-3213(94)00573-W}{\emph{Nucl. Phys. B}
  {\bf 436} (1995) 721--745}, [\href{http://arxiv.org/abs/hep-th/9408169}{{\tt
  hep-th/9408169}}].
  


\bibitem{Maldacena:1997re}
J.~M. Maldacena, \emph{{The Large N limit of superconformal field theories and
  supergravity}}, \href{http://dx.doi.org/10.1023/A:1026654312961,
  10.1023/A:1026654312961}{\emph{Adv.Theor.Math.Phys.} {\bf 2} (1998)
  231--252}, [\href{http://arxiv.org/abs/hep-th/9711200}{{\tt
  hep-th/9711200}}].

\bibitem{Gubser:1998bc}
S.~S. Gubser, I.~R. Klebanov and A.~M. Polyakov, \emph{{Gauge theory
  correlators from non-critical string theory}},
  \href{http://dx.doi.org/10.1016/S0370-2693(98)00377-3}{\emph{Phys. Lett.}
  {\bf B428} (1998) 105--114}, [\href{http://arxiv.org/abs/hep-th/9802109}{{\tt
  hep-th/9802109}}].

\bibitem{Witten:1998qj}
E.~Witten, \emph{{Anti-de Sitter space and holography}}, {\emph{Adv. Theor.
  Math. Phys.} {\bf 2} (1998) 253--291},
  [\href{http://arxiv.org/abs/hep-th/9802150}{{\tt hep-th/9802150}}].

\bibitem{Levy:1969cr}
M.~Levy and J.~Sucher, \emph{{Eikonal approximation in quantum field theory}},
  \href{http://dx.doi.org/10.1103/PhysRev.186.1656}{\emph{Phys. Rev.} {\bf 186}
  (1969) 1656--1670}.

\bibitem{tHooft:1987vrq}
G.~'t~Hooft, \emph{{Graviton Dominance in Ultrahigh-Energy Scattering}},
  \href{http://dx.doi.org/10.1016/0370-2693(87)90159-6}{\emph{Phys. Lett. B}
  {\bf 198} (1987) 61--63}.

\bibitem{Cornalba:2006xk}
L.~Cornalba, M.~S. Costa, J.~Penedones and R.~Schiappa, \emph{{Eikonal
  Approximation in AdS/CFT: From Shock Waves to Four-Point Functions}},
  \href{http://dx.doi.org/10.1088/1126-6708/2007/08/019}{\emph{JHEP} {\bf 08}
  (2007) 019}, [\href{http://arxiv.org/abs/hep-th/0611122}{{\tt
  hep-th/0611122}}].

\bibitem{Heemskerk:2009pn}
I.~Heemskerk, J.~Penedones, J.~Polchinski and J.~Sully, \emph{{Holography from
  Conformal Field Theory}},
  \href{http://dx.doi.org/10.1088/1126-6708/2009/10/079}{\emph{JHEP} {\bf 10}
  (2009) 079}, [\href{http://arxiv.org/abs/0907.0151}{{\tt 0907.0151}}].

\end{thebibliography}
\providecommand{\href}[2]{#2}\begingroup\raggedright\endgroup

\end{document}